\documentclass[%
 reprint,
 amsmath,amssymb,
 aps,
prl,
]{revtex4-2}

\usepackage{graphicx}% Include figure files
\usepackage{dcolumn}% Align table columns on decimal point
\usepackage{bm}% bold math
\begin{document}

\preprint{APS/123-QED}

\title{Topological quantum hodographs}

\author{Nikolay Rosanov} 

\affiliation{Ioffe Institute, Politekhnicheskaya str. 26, St. Petersburg 194021, Russia}

\author{Sergey Fedorov} 

\affiliation{Ioffe Institute, Politekhnicheskaya str. 26, St. Petersburg 194021, Russia}

\author{Mikhail Arkhipov} 

\affiliation{Ioffe Institute, Politekhnicheskaya str. 26, St. Petersburg 194021, Russia}

%\date{\today}% It is always \today, today,
             %  but any date may be explicitly specified

\begin{abstract}
In quantum mechanics, the wave function encodes all information about a particle through its probability density and phase. While stationary states are characterized by conserved quantum numbers such as energy and, in central potentials, angular momentum, non-stationary superpositions --- particularly in anisotropic or time-dependent fields --- generally lack equally universal descriptors of their spatiotemporal dynamics. Here we introduce quantum hodographs: the trajectories traced in time by the expectation value of observable vector quantities.
%the probability current and the Ehrenfest trajectory.
%⟨r(t)⟩ (or dipole moment ⟨d(t)⟩).

For a free electron in a superposition of three plane waves, all hodographs of the probability current lie on a universal cubic surface with conical singularities. Rational frequency (energy) difference ratios produce non-contractible loops with well-defined winding numbers. In anisotropic harmonic oscillators the Ehrenfest trajectories --- hodographs of the expectation value of particle radius-vector --- form three-dimensional Lissajous knots, echoing the classical Thomson vortex-atom model. Externally driven quantum systems allow controllable initiation of knotted hodographs. We propose an optical modulation spectroscopy scheme for reconstructing these topological features in trapped ions and single-electron systems. The topological indices of the loops and knots are robust to parameter variations, offering a new tool for characterizing complex quantum dynamics.
\end{abstract}

%\keywords{Suggested keywords}
%Use showkeys class option if keyword display desired

\maketitle

%\tableofcontents
%\section{Introduction}
In quantum mechanics, the wave function $\Psi(\mathbf{r},t)$ where $\mathbf{r}$ is radius-vector and $t$ is time, encodes complete information about a particle through its probability density 
$|\Psi|^2$ and phase. For stationary states, the eigenfunctions are naturally classified by conserved quantum numbers --- energy and, in spherically symmetric potentials, angular momentum \cite{LandauLifshitz}. However, non-stationary superpositions, especially in anisotropic or time-dependent fields, generally lack equally universal descriptors of their spatiotemporal dynamics.

The evolution of expectation values of $\mathbf{r}$ follows the Ehrenfest theorem, yielding classical-like equations of motion \cite{Ehrenfest}. Yet this description does not capture the full geometric and topological structure of the trajectories traced by $ \langle\mathbf{r}(t)\rangle $ and
$ \langle\mathbf{j}(t)\rangle $ --- the average values of probability current. Recent advances in attosecond physics \cite{Nobel,Krausz,Cruz,Biegert} make it possible to track both the probability density and phase of the electron wave function \cite{Villeneuve}.
%with high resolution.

Here we introduce \textit{quantum hodographs} --- trajectories formed over time by the expected values of various observable vector quantities, namely the probability current $ \langle\mathbf{j}(t)\rangle $, the particle radius vector $ \langle\mathbf{r}(t)\rangle $ (coincides with the Ehrenfest trajectory), and the dipole moment $ \langle\mathbf{d}(t)\rangle $. 
This framework provides a natural and topologically rich characterization of quantum motion where conventional quantum numbers are insufficient. It differs radically from previous topological analyses based on nodal lines --- the zeros of the wave function \(\Psi(\mathbf{r},t)\) \cite{Berry,Enciso,Taylor,Canzani}. On the other hand, hodographs of classical electromagnetic fields have recently been considered, revealing temporal topological knots for polychromatic waves \cite{Sugic,Ferrer} and knots and M\"obius strips for pulses \cite{PRA2025}.

For a free electron in a superposition of three plane waves, all hodographs of  $ \langle\mathbf{j}(t)\rangle $ lie on a \textit{universal cubic surface} with conical singularities. Rational frequency (energy) difference ratios yield non-contractible loops with well-defined winding numbers. In anisotropic harmonic oscillators, Ehrenfest trajectories form Lissajous knots, extending the classical Thomson vortex-atom model \cite{Thomson}. External driving enables controllable initiation of knotted hodographs. We also propose an optical modulation spectroscopy scheme for their reconstruction in trapped-ion experiments.

The starting point is the Schrödinger equation for the wave function \(\Psi\) of an electron (and similar of an ion) subjected to radiation:
\begin{equation}
i \hbar \frac{\partial \Psi}{\partial t} = (\hat{H}_0 + \hat{V}) \Psi.
\label{eq:tdse}
\end{equation}
Here 
%\(t\) is time, 
\(\hbar\) is the reduced Planck constant,
$\hat{H}_0 = -\frac{\hbar^2}{2 m_e} \nabla^2 + U(\mathbf{r})$
--- the unperturbed Hamiltonian, \(U(\mathbf{r})\) is the stationary potential, 
%\(\mathbf{r}\) is the radius vector, 
and $\hat{V}(\mathbf{r}, t) = e \, (\mathbf{r} \cdot \mathbf{E}(\mathbf{r}, t)), \quad e > 0$
--- the interaction potential of the electron with charge \(-e\) and mass \(m_e\) with the external electric field \(\mathbf{E}(\mathbf{r}, t)\) in the dipole approximation.

Next, we will consider the dynamics of the average position \(\langle \mathbf{r} \rangle\) of the electron and the average value of the probability current $\langle\mathbf{j} \rangle$ \cite{LandauLifshitz}
\begin{equation}
\langle \mathbf{r} \rangle = \langle \Psi^* | \mathbf{r} | \Psi \rangle, \quad 
\langle\mathbf{j} \rangle = \frac{\hbar}{m_e} \Im (\Psi^* \nabla \Psi).
\label{eq:rj}
\end{equation}
For hydrogen-like atoms, the average dipole moment $\langle \mathbf{d} \rangle = -e \langle \mathbf{r} \rangle $ and
%the average 
electric current $\langle \mathbf{j_e} \rangle = -e \langle \mathbf{j} \rangle$.
%The quantity \(\langle \mathbf{d} \rangle\) has a clear physical meaning also for such mesoscopic %objects as quantum dots, which are likewise subject to consideration. 

% Free electron
We will begin with the superposition of monoenergetic states of the free electron \(\Psi_n\) with different energies \(E_n\), such that \(\hat{H}_0 \Psi_n = E_n \Psi_n\)
\begin{equation}
\Psi(\mathbf{r}, t) = \sum_{n=1}^{3} C_n \Psi_n(\mathbf{r}, t).
\label{eq:PsiCPsi}
\end{equation}
We assume that these states possess a definite momentum \(\mathbf{p}_n\), 
$\Psi_n = \exp(i \mathbf{k}_n \cdot \mathbf{r} - i \omega_n t),$
where 
$\omega_n = {E_n}/{\hbar} = {p_n^2}/{(2m_e\hbar)}$ and $\mathbf{k}_n = {\mathbf{p}_n}/{\hbar}$. 
%\begin{equation}
%\omega_n = \frac{E_n}{\hbar} = \frac{p_n^2}{2m_e\hbar},
%\quad
%\mathbf{k}_n = \frac{\mathbf{p}_n}{\hbar}.
%\label{eq:wk}
%\end{equation}
All energy values (frequencies \(\omega_n\)), momenta \(\mathbf{p}_n = p_n \mathbf{e}_n\) and wave vectors \(\mathbf{k}_n = k_n \mathbf{e}_n\) differ. The unit vectors \(\mathbf{e}_n\) are taken to be Cartesian basis vectors.

The components of the probability current, Eqs.~(\ref{eq:rj}), for such superposition
%in Eq.~(\ref{eq:PsiCPsi})
%with Eqs.~(\ref{eq:wk}) 
are
\begin{equation}
\label{eq:J3waves}
\begin{aligned}
\frac{m_e j_1}{\hbar k_1} &= |C_1 C_2| \cos(\omega_{21} t + \Phi_{21}) + |C_1 C_3| \cos(\omega_{31} t + \Phi_{31}), \\
\frac{m_e j_2}{\hbar k_2} &= |C_2 C_1| \cos(\omega_{21} t + \Phi_{21}) + |C_2 C_3| \cos(\omega_{32} t + \Phi_{32}), \\
\frac{m_e j_3}{\hbar k_3} &= |C_3 C_1| \cos(\omega_{31} t + \Phi_{31}) + |C_3 C_2| \cos(\omega_{32} t + \Phi_{32}).
\end{aligned}
\end{equation}
Here \(\omega_{ln} = \omega_l - \omega_n\),
$\Phi_{ln} = - (\arg C_l - \arg C_n) - (k_l x_l - k_n x_n)$,
and the time-independent term is omitted, which only shifts the “center" of the hodograph. The probability current oscillates with time at three (difference) frequencies, with \(\omega_{31} = \omega_{21} + \omega_{32}\) and \(\Phi_{31} = \Phi_{21} + \Phi_{32}\)
--- the frequency and phase matching conditions. 
For three waves the hodograph is certainly non-planar, which manifests itself in the geometric phase of Aharonov and Anandan~\cite{Aharonov1987} differing from \(0\) and \(\pi\). Naturally, the hodograph changes with frequency variations. Nevertheless, there exists a surface on which the hodographs for all admissible frequencies lie. It is obtained by the standard elimination of time from Eqs.~(\ref{eq:J3waves}), see, e.g.,~\cite{Jones}, and by a linear transformation of the current components
\begin{equation}
\label{eq:cosJ}
\begin{aligned}
\cos(\omega_{21} t + \Phi_{21}) &= 
\frac{m_e}{2\hbar |C_1 C_2|}
\left( \frac{j_1}{k_1} + \frac{j_2}{k_2} - \frac{j_3}{k_3} \right)
\equiv J_1, \\
\cos(\omega_{32} t + \Phi_{32}) &= 
\frac{m_e}{2\hbar |C_2 C_3|}
\left( -\frac{j_1}{k_1} + \frac{j_2}{k_2} + \frac{j_3}{k_3} \right)
\equiv J_2, \\
\cos(\omega_{31} t + \Phi_{31}) &= 
\frac{m_e}{2\hbar |C_3 C_1|}
\left( \frac{j_1}{k_1} - \frac{j_2}{k_2} + \frac{j_3}{k_3} \right)
\equiv J_3.
\end{aligned}
\end{equation}
As a result, we get
%arrive at 
the universal a surface with algebraic equation of a third-degree independent of the wave parameters
\begin{equation}
F(J_1, J_2, J_3) = J_1^2 + J_2^2 + J_3^2 - 2 J_1 J_2 J_3 - 1 = 0,
\quad |J_i| \le 1.
\label{eq:F_J}
\end{equation}

\begin{figure}[ht]
\centering
\includegraphics[width=0.45\textwidth]{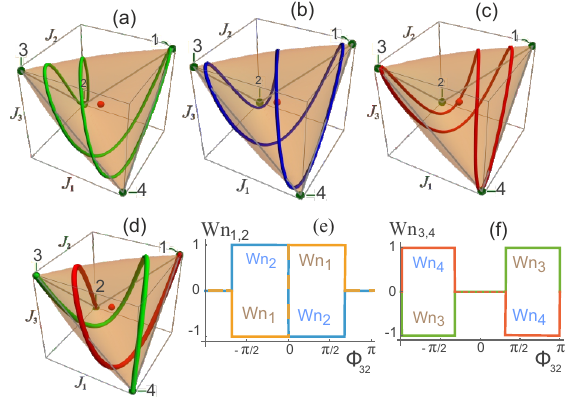}
\caption{(Color online) (a-c) Universal surface, Eq.~(\ref{eq:F_J}), with four conic points 1--4, %Eq.~(\ref{eq:Cone}), 
and hodographs of the probability current $\mathbf{J}(t)$, for three waves, Eq.~(\ref{eq:J3waves}), for $\omega_{32}=2\omega_{21}$ and phase $\Phi_{21}=\Phi_{32}=\pi/4$ (a), 
$\pi/2$ (b), $3\pi/4$ (c), 0 (d, green line) and $\pi$ (d, red line). 
Red points mark the origin of coordinates.
%$J_n$. 
(e,f) Dependence of topological index $\mathrm{Wn}$ --- winding number relative the conical points --- on the phases.
%$\Phi_{21}=\Phi_{32}$.
}
\label{Fig:Loops}
\end{figure}

Fig.~\ref{Fig:Loops}a--d shows this closed surface of the hodographs.
%is shown in Fig.~\ref{Fig:Loops}a--d. 
Important 
%for the subsequent analysis 
are its four 
%special 
conical points of the second order: (1) (1,1,1), (2) (-1,1,-1), (3) (-1,-1,1), and (4) (1,-1,-1).
%\begin{subequations}\label{eq:14}
%\begin{equation}\label{eq:Cone}
%\begin{aligned}
%(1) \quad (1,1,1), \quad (2) \quad (-1,1,-1), \\
%(3) \quad (-1,-1,1), \quad (4) \quad (1,-1,-1).
%\end{aligned}
%\end{equation}
These points are determined by the condition 
$\partial F/\partial J_n =0$, $n=1,2,3.$
%$\partial F/\partial J_1 = \partial F/\partial J_2 = \partial F/\partial J_3 = 0$. 
In Eqs.~(\ref{eq:cosJ}) they correspond to zero frequencies $\omega_{mn}=0$ and phases $\Phi_{mn}=0$ or $\pi$~(static fields). Near these points the surface has the form of cones.

Eqs.~(\ref{eq:cosJ}) define the hodographs as Lissajous knots \cite{Lis,Lis2,Lis3}. The presence of surface~(\ref{eq:F_J}) precludes the possibility of nontrivial knots. 
When the beat frequencies and phase differences are fixed, the motion along the hodograph with time is quasiperiodic in the general case. For an irrational frequency ratio the hodograph gradually fills the entire surface.
% (Fig.~\ref{Fig03}) Suppl. Mat.    ??
However, if the ratio of the beat frequencies \(\omega_{21}\) and \(\omega_{32}\) is rational, the motion is periodic and the hodographes are loops hooked onto two conical points; moreover, the loops are non-contractible to a point on the surface. They are characterized by the topological index $\mathrm{Wn}$, the winding number, equal to the difference between the number of clockwise and counterclockwise revolutions around a conical point during one period of the hodograph traversal. There exist two special loops degenerated into curves that connect pairs of conical points (Fig.~\ref{Fig:Loops}d); the conical points are cusps for them. When the phase difference changes, the loops smoothly pass from one special loop to another (Fig.~\ref{Fig:Loops}a--c and Animation 1 in \cite{supp_mat}). The type of loop---i.e., its attachment to one and the same pair of conical points, and the winding numbers, Fig.~\ref{Fig:Loops}e,f, are preserved under a significant change of phases.

\begin{figure}[ht]
\centering
\includegraphics[width=0.45\textwidth]{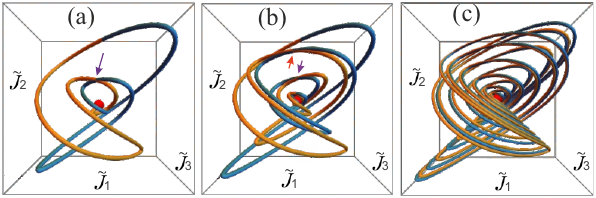}
\caption{(Color online) Hodographs of the probability current for $\omega_{32}=2\omega_{21}$, $\Phi_{21}=1.5$, $\Phi_{31}=0.2$, $\Phi_{31}=1.7$, and $\alpha=1$ (a), 0.25 (b), 0.01 (c).  
\label{Fig:PulseMebius}
}
\end{figure}

Wave packets with a finite temporal duration are of practical importance. In this case, the hodographs are always closed, since the probability of finding the electron in the region vanishes both before and after the passage of the packets; we disregard the purely mathematical singularity in the limit $t \to \pm \infty$. Although the hodographs no longer lie on the universal surface, Eq.~(\ref{eq:F_J}), the 
%frequency and phase 
matching conditions 
%$\omega_{31} = \omega_{21} + \omega_{32}$ and $\Phi_{31} = \Phi_{21} + 
%\Phi_{32}$ 
still preclude the formation of nontrivial knots; only unknotted loops are present. The finite duration of the packets can be modeled by introducing a Gaussian temporal envelope: 
$ \tilde{J}_n = J_n \exp(-\alpha t^2)$,
where $J_n$ are given by Eqs.~(\ref{eq:cosJ}). For $\alpha = 0$ we recover the previous case. However, as $\alpha \to 0^+$, the number of hodograph crossings in its planar projections 
%$\mathrm{Cr}$ 
increases to infinity, with crossings accumulating at the fronts $t \to \pm \infty$ (see Fig.~\ref{Fig:PulseMebius}a--c).

Next, we illustrate hodographs for bound states by the example of a 3D anisotropic harmonic oscillator. For it, the potential in 
%the Schrödinger equation
Eq.~(\ref{eq:tdse}) is
$U(\mathbf{r}) = \frac{m_e}{2} \sum_{n=1}^3 \omega_n^2 x_n^2$ where 
%the quantities 
\(\omega_n\) represent the frequencies of harmonic oscillations of the corresponding classical oscillator.
%Such a potential is realized with high accuracy in Penning 
%traps 
%or Paul traps for electrons and ions~\cite{dehmelt,Brown,Blaum,Matthiesen}.

The Schrödinger equation with such potential allows the separation of variables:
$\Psi = \prod_{n=1}^{3} \psi_n(x_n,t)$.
Thus, we arrive at the equations of the one-dimensional harmonic oscillator
\begin{equation}
i\hbar\frac{\partial\psi_n}{\partial t}=-\frac{\hbar^2}{2m_e}\frac{\partial^2\psi_n}{\partial x_n^2}
+\frac{m_e\omega_n^2 x_n^2}{2}+eE_n x_n\,\psi_n.
\label{eq:Osil1}
\end{equation}
The solution of Eq.~(\ref{eq:Osil1}) for an arbitrary dependence \(E_n(t)\) is expressed, up to a factor inessential for determining the mean value of the coordinate and the dipole moment, through the eigenfunctions \(\Psi_n\) of the Hamiltonian \(\hat{H}_0\)~\cite{BZP}:
\begin{equation}
\psi_n(x,t)=\Psi_n\bigl[x-x_{\rm cl,n}(t)\bigr].
\label{eq:Osil2}
\end{equation}
Here \(x_{\rm cl,n}(t)\) is the solution of the equation of motion of the classical oscillator
\begin{equation}
\frac{d^2x_{\rm cl,n}}{dt^2}+\omega_n^2 x_{\rm cl,n}+eE_n=0.
\label{eq:Osil3}
\end{equation}
It follows that the hodograph of the mean value of the electron coordinates coincides with its classical trajectory,
$\langle x_n\rangle=x_{\rm cl,n}(t),\qquad\langle d_n\rangle=-e\langle x_n\rangle$.
This corresponds to the Ehrenfest theorem~\cite{Ehrenfest}.

In the case of not driven electrons (in the absence of an external field, \(E_n(t)=0\)), we obtain the following.
\begin{equation}
\langle x_n\rangle=a_n\cos(\omega_n t+\Phi_n),\qquad\langle d_n\rangle=-ea_n\cos(\omega_n t+\Phi_n).
\label{eq:Osil4}
\end{equation}
These dependencies are close to Eqs.~(\ref{eq:cosJ}), but now the matching conditions are not applicable. These are again Lissajous knots; however, the universal surface of the hodographs may be absent. Accordingly, periodically repeating topologically nontrivial knots become possible if the frequency ratios are rational numbers. These knots naturally reproduce reproduce the model of ``vortex atoms'' of Thomson~\cite{Thomson}. 

\begin{figure}[ht]
\centering
\includegraphics[width=0.45\textwidth]{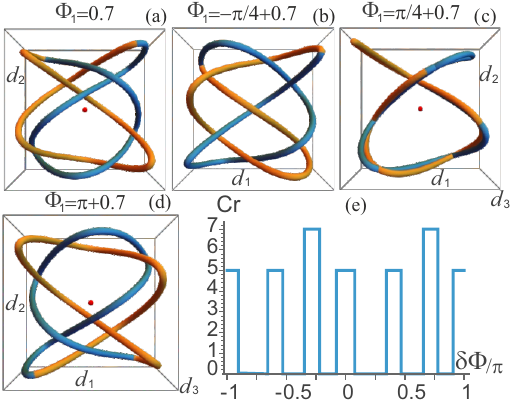}
\caption{(Color online)
Transformations of the 3-twist knot $5_2$. (a–d): Projections of the knot onto the $xy$-plane at various phase shifts  $\delta \Phi$. 
(e): Dependence of 
%the crossing number 
$\mathrm{Cr}$ on $\delta \Phi$; $a_n = 1$.
}
\label{Fig:3twist}
\end{figure}

Fig.~\ref{Fig:3twist} and Animation 2 in \cite{supp_mat}
show transformation of the initial Lissajous knot \(5_2\) (the 3-twist knot, Fig.~\ref{Fig:3twist}a). For it, $ \omega_1 : \omega_2 : \omega_3 = 3 : 2 : 7$ and phases 
$\Phi_1 = 0.7 + \delta \Phi$, $\Phi_2 = 0.2 + \delta \Phi$,  $\Phi_3 = 0$ with various phase shifts $\delta \Phi$ in Eq.~(\ref{eq:Osil4}). 
As seen in Fig.~\ref{Fig:3twist}e, the crossing number, the topological invariant $\mathrm{Cr}$, i.e., the minimum number of crossings over any plane projections, is preserved within some finite intervals of phases and varies from one interval to another. Correspondingly, the knot type varies. In particular, 
Fig.~\ref{Fig:3twist}b shows the prime arborescent knot $7_4$,
%with 7 crossings
and intervals with $\mathrm{Cr}=0$ correspond to unknots.
In view of Eq.~(\ref{eq:Osil4}), Fig.~\ref{Fig:3twist} can be interpreted as referring both to Ehrenfest trajectories and to hodographs of the dipole moment.
%$\langle x\rangle=a_x\cos(2\omega_0 t+0.2),\qquad\langle y\rangle=a_y\cos(3\omega_0 t+0.7)$
%$\langle z\rangle=a_z\cos(7\omega_0 t)$,
%the Aharonov--Anandan phase \(\gamma_{AA}\approx0.95\pi\). Lissajous knots do not change their type under arbitrary changes in the amplitudes \(a_n\) (they must remain nonzero). As seen from Fig.~\ref{Fig:3twist}e, %the topological index Writhe is preserved under comparatively large changes in the phases \(\Phi_n\).

Next, for polychromatic radiation,
$\mathbf{E} = \sum_{n=1}^{3} A_n \mathbf{e}_n \cos(\Omega_n t + \varphi_n)$,
the steady-state solutions of Eq.~(\ref{eq:Osil3}) yield the mean dipole moment
\begin{equation}
\langle \mathbf{d} \rangle = \sum_{n=1}^{3} A_n \alpha_n \cos(\Omega_n t + \varphi_n) \mathbf{e}_n . \label{eq:ad}
\end{equation}
The dynamic susceptibility
%$\alpha_n = \frac{e^2}{m_e (\omega_n^2 - \Omega_n^2)}$
$\alpha_n = {e^2}/{(m_e (\omega_n^2 - \Omega_n^2))}$
exhibits sharp resonances. Consequently, knotted hodographs of the mean dipole moment become possible even for weak external radiation when the frequencies \(\Omega_n\) of the external radiation are in rational ratios. They are described by the same Eq.~(\ref{eq:Osil4}) with replacement $\omega_n \to \Omega_n$. 

Initiation of nontrivial Ehrenfest trajectories is possible by means of pulsed external 
radiation with finite duration. After the pulse $x_{{\rm cl},n}(t) = -({e}/{\omega_n}) \left[ c_n \sin(\omega_n t) - s_n \cos(\omega_n t) \right]$
%\begin{equation}
%x_{{\rm cl},n}(t) = -\frac{e}{\omega_n} \left[ c_n \sin(\omega_n t) - s_n \cos(\omega_n t) \right]
%\label{eq:x_cl_n}
%\end{equation}
where 
%$c_n \approx \int_{0}^{\infty} E_n(\tau) \cos(\omega_n \tau) \, d\tau$
$c_n \approx \int E_n(\tau) \cos(\omega_n \tau) \, d\tau$
and 
$s_n \approx \int E_n(\tau) \sin(\omega_n \tau) \, d\tau$.
Thus, after the end of the pulse, harmonic oscillations continue at the frequency 
$\omega_n$. Their phase is determined by the shape of the pulse $E_n(t)$.

Equation~(\ref{eq:ad}),
for various forms of dynamic susceptibility $\alpha_n$,
holds for a broad class of 
%quantum 
micro-objects, including quantum dots, and for classical dipoles.
%The 
%key distinction is that 
%quantum wave-packet blurring is absent 
%only 
%for a harmonic potential. Otherwise, for finite-depth potentials, the %given expressions cannot be used for highly excited states, where %ionization processes become important.
The formation of knotted Ehrenfest trajectories is not limited to harmonic traps. In any confining potential possessing at least two bound states, a coherent superposition of these states leads to periodic or quasiperiodic oscillations of the expectation values 
$\langle x_n (t)\rangle$ 
at the transition frequencies. 
When multiple transition frequencies are commensurate, topological
%Lissajous 
knots remain possible. In all these cases, reverse probability flows occur
that has been widely discussed for 1D quantum systems~\cite{Allcock1969,Miller2021}.
%, provided the dynamics stays within the few-level subspace on timescales short compared to decoherence and multi-level dephasing."
%Topologically non-trivial knots with suppressed wave packet blurring are achieved not only for a harmonic potential, but also for an arbitrary potential that allows separation of variables and has at least two stationary states of the discrete spectrum. Indeed, in a superposition of two levels, the average value of each Cartesian coordinate varies sinusoidally with time, with a frequency corresponding to the energy difference between these levels. Thus, with a rational ratio of beat frequencies, various Lissajous knots are still possible.

Motion of classical \cite{BornWolf} and quantum \cite{Scully1997} dipoles $\mathbf{d}(t)$ generates radiation. Near-field radiation hodographs have the same topology as dipoles \cite{supp_mat}). In the far zone, the electromagnetic field becomes transverse, nontrivial knots are impossible there. Figure S4 and Animation 3 in \cite{supp_mat} illustrate the transition from near- to far zone.

%with electric field
$\mathbf{E}_s = \sum_{n=1}^{3} \mathbf{E}_n$, where~
%\cite{BornWolf}
\begin{equation}
\begin{split}
\mathbf{E}_s(\mathbf{r},t) &= 
\left\{
\frac{3(\mathbf{d}\cdot\mathbf{r})}{r^5} 
+ \frac{3(\dot{\mathbf{d}}\cdot\mathbf{r})}{c r^4} 
+ \frac{(\ddot{\mathbf{d}}\cdot\mathbf{r})}{c^2 r^3}
\right\} \mathbf{r} \\
&\quad - 
\left(
\frac{\mathbf{d}}{r^3} 
+ \frac{\dot{\mathbf{d}}}{c r^2} 
+ \frac{\ddot{\mathbf{d}}}{c^2 r}
\right).
\label{eq:BoWo}
\end{split}
\end{equation}

The topological features of quantum hodographs predicted in this work are within reach of current experimental capabilities, particularly in trapped-ion and single-electron systems. Paul and Penning traps already provide high-accuracy harmonic or near-harmonic confinement \cite{dehmelt,Brown,Blaum,Matthiesen}, while modern techniques allow coherent control and readout of motional states.

A practical detection scheme is as follows. The system is driven into a coherent forced-oscillation regime by three monochromatic microwave (or terahertz) fields propagating along mutually orthogonal directions. Such fields can be generated by quantum cascade lasers \cite{Belkin2008Room,Piccardo2018}. Choosing commensurate frequencies (for example, in the ratio 2:3:5) induces complex three-dimensional Lissajous motion of the expectation value $ \langle\mathbf{r}(t)\rangle $.
%In this driven steady state the Ehrenfest theorem remains strictly valid.

A linearly polarized monochromatic optical probe beam (visible or near-infrared) is passed through the trap at an optimized angle --- preferably close to the body diagonal relative to the microwave axes. The oscillatory motion imprints multi-tone phase and amplitude modulation onto the transmitted probe field. Because relative phases between the modulation components are preserved upon propagation, high-speed digitization of the transmitted intensity or quadrature signals, followed by multi-frequency digital demodulation or Fourier analysis, enables extraction of both amplitudes and relative phases of all three orthogonal components.

Importantly, only the three-frequency non-coplanar driving configuration provides access to the full spatial structure and topology of the quantum motion. While one or two frequencies reveal only the presence of oscillation or planar motion, three commensurate frequencies generate genuinely 3D-Lissajous figures, including nontrivial knots with well-defined topological indices.

This approach enables experimental reconstruction of the complete Ehrenfest trajectory 
$ \langle\mathbf{r}(t)\rangle $ in a driven quantum superposition and offers a direct platform for verifying the Ehrenfest theorem in three dimensions under nontrivial driven dynamics --- a test that has remained elusive with conventional techniques. Although the lifetime of topological features is limited by decoherence (trap fluctuations, spontaneous emission), modern trapped-ion systems can maintain coherence for milliseconds or longer in the driven regime, sufficient to record multiple periods of the hodograph and extract its invariants.

Concluding, we have introduced quantum hodographs --- not only the Ehrenfest trajectories $ \langle\mathbf{r}(t)\rangle $, but also the trajectories traced by the expectation values of the probability current $ \langle\mathbf{j}(t)\rangle $ and the dipole moment $ \langle\mathbf{d}(t)\rangle $ --- and analyzed their three-dimensional topology. For a free electron in a superposition of three plane waves, all hodographs lie on a universal cubic surface with conical singularities. Rational frequency difference ratios produce non-contractible loops characterized by winding numbers. In anisotropic harmonic oscillators, Ehrenfest trajectories form Lissajous knots. Driven systems allow external initiation of such knotted structures.

We propose an optical modulation spectroscopy scheme using three orthogonal commensurate microwave drives that enables reconstruction of these topological features in trapped ions. This provides a practical way to visualize quantum hodographs and test the Ehrenfest theorem in complex driven dynamics.

The topological indices are robust and offer a new tool for characterizing quantum motion when traditional quantum numbers lose their power. Experimental realization of quantum hodographs will advance time-domain quantum physics and precision control of individual charges.

%\section{Acknowledgments}

%\acknowledgments
%The authors used Grok in order to improve the 
%language and 
%style of the manuscript. 
This work was supported 
% by Ioffe Institute State Assignment, topic 0040-2019-0017 (calculation of quantum hodographs) and 
by the Russian Science Foundation, grant 21-72-30020-$\Pi$.

\bibliography{article}% Produces the bibliography via BibTeX.

\end{document}